# Avrami exponent under transient and heterogeneous nucleation transformation conditions

I. Sinha[a,*] and R. K. Mandal[b]

[a]Department of Applied Chemistry, Institute of Technology,
Banaras Hindu University, Varanasi 221005, India

[b]Centre of Advanced Study, Department of Metallurgical Engineering,
Institute of Technology, Banaras Hindu University, Varanasi 221005, India

[*]Corresponding author
[*]Email: isinha.apc@itbhu.ac.in,
Tel. 91-542-2454072 Fax: +91 542 2368428.

**Abstract**

The Kolmogorov-Johnson-Mehl-Avrami model for isothermal transformation kinetics is universal under specific assumptions. However, the experimental Avrami exponent deviates from the universal value. In this context, we study the effect of transient heterogeneous nucleation on the Avrami exponent for bulk materials and also for transformations leading to nanostructured materials. All transformations are assumed to be polymorphic. A discrete version of the KJMA model is modified for this purpose. Scaling relations for transformations under different conditions are reported.

## 1. Introduction

An important problem pertaining to phase transformation relates to deciphering its mechanism from the volume fraction of material transformed $X(t)$ as a function of time. The Kolmogorov, Johnson, Mehl and Avrami (KJMA) theory [1-5] for isothermal kinetics has been used extensively for deducing the mechanism of phase transformations that occur via nucleation and growth. This theory is universal if the assumed conditions in the model are not violated [6]. However, such conditions are seldom realized experimentally [7].

In the KJMA model, transformation is initiated by homogeneous nucleation with the nuclei assumed to have negligible initial radius $(R_c)$. The transformation is supposed to occur in an infinite medium, that is, the linear size of the system $(L)$ is much larger than the distance between the nuclei $(\xi)$ of the new phase. The stable nuclei of the product phase are considered to grow isotropically at a constant rate $(\gamma)$. Growth stops at points of impingement and continues unabated elsewhere. Under these assumptions the KJMA model expresses $X(t)$ in terms of extended volume fraction transformed $X_{ex}(t)$ as the following.

$$X = 1 - \exp(-X_{ex}) \qquad (1)$$

Here, $X_{ex}$ is the volume fraction transformed if all grains were assumed to grow unimpeded. $X_{ex}(t)$ is typically given as $X_{ex}(t) = kt^n$ where $k$ is a constant, $n$ is an number and $t$ is time elapsed since the start of the transformation [8]. For example, under constant nucleation rate $I$ and three-dimensional (3D) spherical growth, $k = I\frac{\pi}{3}\gamma^3$ and $n$ = 4. The time exponent ($n = 4$ in this case) or Avrami exponent is found from the slope of the line produced by plotting the experimental values of ln(ln(1/(1- $X$ ))) against ln($t$). A system is defined by its characteristic length, $\xi = \left(\gamma/I\right)^{1/D+1}$ where $I$ and $\gamma$ denote constant nucleation and growth rate respectively [6]. Irrespective of the value of $\xi$, the time exponent value is universal for a system of dimensionality $D$.

In recent publications [9, 10], we have demonstrated that transformations initiated by nuclei of finite size ( $R_c < \xi$ ) give rise to negative deviations in Avrami time exponent $(n)$ from the universal value. The condition of finite nuclei size was considered for bulk polymorphic nanocrystallization transformations. The exponent tends to the universal value of 4 (for transformation of a 3D system) by a linear scaling relation as the condition $R_c < \xi$ tends to $R_c \ll \xi$ [10]. We surmise that other conditions leading to violation of the KJMA assumptions may also display such scaling relations. In the present communication, we take into account the effects of transient nucleation and heterogeneous nucleation on the Avrami exponent. Only polymorphic transformations are studied. Such transformations are frequently observed during bulk crystallization of metallic glasses [11-15]. We also consider the effects of these violations on the linear scaling relations that occur due to finite size nuclei. Clavaguera-Mora et al. have reviewed various other aspects of such kinetics [16].

Transient nucleation and heterogeneous nucleation conditions give rise to different Avrami exponent values. Thus we obtain $n > 4$ for transient nucleation and $n$ values between 3 and 4 owing to heterogeneous nucleation [8]. Further, as already known [9], the exponent for bulk nanocrystallization shows negative deviation from the universal value. At present, when we have a combination of the above-mentioned factors, then it becomes difficult to predict a nucleation mechanism from the KJMA exponent. Our objective in this paper is to find relations that will enable us to identify the underlying mechanism of polymorphic transformation under such combinations of conditions.

Let us consider a liquid that is quenched rapidly from above its melting point to a temperature T below its melting point. The distribution of crystal embryos changes with time to reflect the changed conditions. Consequently, the nucleation frequency is small at the beginning of the isothermal hold and increases until the steady state distribution is established. The transient nucleation time or the time required by the transforming system to achieve constant nucleation rate is, therefore, dependent on the shape of underlying cluster size distribution [17, 18]. The time required to achieve the steady state nucleation rate is called transient time $\tau$. The time-dependent isothermal nucleation frequency $I(t)$ is usually expressed by the following approximate relation [19] by Zeldovich:

$$I(t) = I_s \exp\left(-\tau/t\right) \tag{2a}$$

where $I_s$ is the steady-state nucleation frequency. The Zeldovich expression is suitable for the simple polymorphic transformation condition that we focus on in this paper [20]. Initially, we consider the effects of different transient nucleation times ($\tau$) on Avrami exponents for transformations via homogeneous nucleation on systems defined by different characteristic lengths. The effects of finite sized nuclei on such transformation kinetics, relevant for bulk nanocrystallization kinetics, are studied next.

The isothermal crystallization kinetics is analyzed after the following modified Avrami expression.

$$X(t) = 1 - \exp\{-k(t - t_{inc})^n\} \tag{2b}$$

Here $X(t)$ is the crystallized volume fraction, $t$ the annealing time, $t_{inc}$ the incubation time, $n$ a exponent related to the dimensionality of nucleation and growth, and $k$ a reaction rate constant. The exponent $n$ is determined using the form given below.

$$\ln[-\ln(1 - x)] = \ln k + n \ln(t - t_{inc}) \tag{2c}$$

The incubation time $(t_{inc})$ is the time interval between the specimen reaching the annealing temperature and the time at which observable transformation occurs. Increase in $\tau$ delays the achievement of the steady state or constant nucleation rate. Since nucleation represents the start of the transformation, therefore, such transient nucleation delays the KJMA kinetics by $t_{inc}$. Thus, transient nucleation shifts the time origin and is proportional to the incubation time [21].

Assuming transformation by isothermal annealing at the peak transformation temperature $(T_{\mathrm{P}})$, our study on the effect of different transient nucleation times on the Avrami exponent may represent two experimental situations. In the first case, keeping the differential scanning calorimeter (DSC) heating rate fixed, we consider isothermal annealing at temperatures approaching $T_{\mathrm{P}}$. The incubation time is found to be the least when heat treatment is carried out at the peak transformation temperature and it increases [22, 23], as one moves away from $T_P$. Thus, transient nucleation time increases as the annealing temperature moves away from $T_P$. The other experimental case arises when the DSC heating rate is varied. For faster heating rates the transformation is sharply peaked around the nucleation temperature. Incubation time increases for slower heating rates. This indicates the system has access to cluster size distributions at temperatures less than $T_P$ and slower heating rates would lead to increase in transient nucleation times [24-26].

The classical theory of homogeneous nucleation deals with phase transformation initiated by nuclei of spherical shape. In heterogeneous nucleation process, nuclei are formed on the surface of a foreign particle (e.g. grain refiner or mould wall) and are a part of a sphere. In the present model we assume that nucleants (grain refiners) are randomly distributed throughout the volume of the parent system. The number of such nucleants is limited and is expected to be less than the upper limit given by homogeneous nucleation. Nucleation occurs on the surface of the nucleants with $\theta(180^\circ < \theta < 0^\circ)$ as the contact angle between the product phase and the surface [27]. In case the nuclei formed are finite in comparison to the characteristic length ($\xi$), the shapes of such finite nuclei do affect the Avrami exponent. For a given $\theta$ value, the Avrami exponent follows a linear scaling relation with decreasing finite nuclei size [10]. Here we investigate the case when $\theta$ is increased for transformation by nuclei having a fixed finite size. Of course, experimentally, it is not possible to determine independently the Avrami exponents of the nuclei as a function of $\theta$. We, however, are considering the effect of predominant $\theta$ that governs the kinetics of transformation. Subsequently, we also try to understand the complications owing to addition of the transient nucleation conditions to the above.

## 2. The Discrete Model

*2.1 Transient homogeneous nucleation*

In case of transformation by homogeneous nucleation we use the phantom nuclei concept [6-8]. The critical volume of each of the spherical nuclei with radius $R_c$ is $v_\tau = \frac{4}{3}\pi R_c^3$ (in $L^3$ units) for homogeneous nucleation. We shall indicate progress of time by an iteration index $i$. We introduce $V_{ur}$, $V_{ac}$ and $V_{ex}$ respectively as untransformed, actual transformed and extended volumes. At $t = 0$ (corresponding to $i = 0$), these are $V_{ur}(0) = V_0$, $V_{ac}(0) = 0$ and $V_{ex}(0) = 0$, where $V_0$ is the initial volume of the parent phase. We recast equation (2), for an iteration interval $(i \geq 1)$ as $N(i) = N_s \exp(\tau/i)$ (where $N_s$ and *N(i)* are the number of nuclei born per unit time per unit volume corresponding to the steady and the transient nucleation rates) number of nuclei is made available in random distribution of nuclei by their proportional distribution in transformed and untransformed volumes. The value of $N_s$ is kept fixed for all computations. The expression of extended volume in the $i^{\text{th}}$ iteration can be written in a manner analogous to our earlier work [9] and is computed by the following equation.

$$V_{ex}(i) = V_{ex}(i-1) + N(i) v_\tau V_{ur}(i-1) + N(i) v_\tau V_{ac}(i-1) + V_{gr}(i) \quad (3)$$

where $V_{gr}(i) =$ increment in growth in the iteration $i$ of the existing nuclei/grains. The term *N(i)* in the above takes care of the transient effects. The term $V_{gr}(i)$ is computed using the following expression,

$$V_{gr}(i) = V_{gr}^{ur}(i) + V_{gr}^{ac}(i) \quad (4a)$$

$V_{gr}^{ur}(i)$ is the increment in growth of nuclei formed on untransformed volume and $V_{gr}^{ac}(i)$ is the increment in growth of the phantom nuclei.

$$V_{gr}^{ur}(i) = \sum_{z=1}^{i-1} N(z) \left[ v_{(i-z)} - v_{(i-(z+1))} \right] . V_{ur}(z-1) \quad (4b)$$

where $i > 1$ and

$$V_{gr}^{ac}(i) = \sum_{z=1}^{i-1} N(z) \left[ v_{(i-z)} - v_{(i-(z+1))} \right] . V_{ac}(z-1) \quad (4c)$$

where $i > 2$. Here $v_i = \frac{4}{3}\pi(R_c + i.\Delta r)^3$ and growth rate $\gamma = \frac{\Delta r}{\Delta t}$. Thus $\Delta r$, is the increment in the radius of each growing nuclei in every iteration interval $\Delta i$. $V_{ac}(i)$ is calculated using the following equation.

$$V_{ac}(i) = V_{ac}(i-1) + N(i).V_{ur}(i-1) v_\tau + V_{gr}(i)\{1 - X(i-1)\} \quad (5)$$

and the fraction transformed is obtained from $X(i) = V_{ac}(i)/V_0$ . Please note that the last term in Eq. (5) represents the contribution of $V_{gr}(i)$ to actual transformed volume. This is proportional to the available untransformed volume.

*2.2 Transient heterogeneous nucleation*

Under this condition the nuclei are formed only in the untransformed volume on nucleants in the parent phase. The corresponding nucleation rate is defined as the number of nuclei per unit time per unit untransformed volume [28]. As mentioned earlier, we assume heterogeneous nucleation is facilitated by the presence of randomly distributed planar nucleants in the volume of the parent phase. Nucleation occurs on the surface of the nucleants with $\theta$ ($180^\circ < \theta < 0^\circ$) as the contact angle between the product phase and the surface. Each nucleant provides a site for only one nucleation event. The following relation gives the volume of the heterogeneous nuclei

$$v_\tau^{het} = v_\tau . S(\theta) \tag{6}$$

where $S(\theta)\left(= 2 - 3\cos\theta + \cos^3\theta/4\right)$ is the shape factor and $v_\tau$ is the critical volume of the homogeneous nucleus. The available number of nucleants and corresponding nucleation events are limited. They may get exhausted before the completion of the transformation. However, in our computations we assume that the number of nucleants is sufficiently large and the phase transformation process is completed before heterogeneous nucleation events are exhausted.

The expression for $V_{ex}$ (i) for heterogeneous transient nucleation takes the form of

$$V_{ex}(i) = V_{ex}(i-1) + N(i).v_\tau^{het} V_{ur}(i-1) + V_{gr}(i) \tag{7}$$

Further, the growth of nuclei is computed assuming their growth to retain the shape till the end of transformation. If we denote $i_{het}$ as the iteration when all nucleants have been utilized, then for the condition $i < i_{het}$ and $i > 1$,

$$V_{gr}(i) = \sum_{z=1}^{i-1} N(z)\left[v_{(i-z)}^{het} - v_{(i-(z+1))}^{het}\right].V_{ur}(z-1) \tag{8a}$$

The corresponding $V_{ac}$ is written as following.

$$V_{ac}(i) = V_{ac}(i-1) + N(i).V_{ur}(i-1)v_\tau^{het} + V_{gr}(i)\{1 - X(i-1)\} \tag{8b}$$

**3. Results**

As mentioned earlier, the incubation time takes into account the time lag before $X$ reaches measurable values. In normal experimental situations, the incubation time $t_{inc}$ is defined as the as the time scale between $t_0$ and $t_{1\%}$, where $t_0$ is the time to reach the annealing temperature and $t_{1\%}$ is the time to reach 1% crystallized volume fraction. On the other hand, at high values of the fraction transformed, the nucleation rate starts to decrease as the nucleation sites become filled with nuclei [29]. Thus, the ends of the Avrami plots deviate from the expected constant nucleation rate straight-line plot. We, therefore, compute and analyze the fraction-transformed data in the range $X = 0.01 - 0.95$, assuming $t_{inc} = t_{1\%}$. The error due to very late stage transformation is avoided by analyzing data only up to $X = 0.95$. Further, it is important to mention that $X(t)$ is itself a functional of the parameter determined directly in the experiment, for example, electrical resistivity. An uncertainty in determination of the experimental observable will cause a new uncertainty in $X(t)$ [29, 30]. The Avrami approach does not concern such experimental noise.

The Avrami exponent $n$ is determined from the slope of the plots of $\ln[-\ln(1 - x)]$ versus $\ln(t - t_{inc})$ assuming increasing transient nucleation times. As described in the Introduction, increasing values of $\tau$ may be assumed to correspond to experimental annealing temperatures at increasing distance from the peak transformation temperature. On the other hand different rates of heating to a given isothermal annealing temperature will also result in different transient nucleation times.

As mentioned earlier, to define the system we need to calculate its characteristic length and time. In comparison to the steady state nucleation case, here we define the characteristic time by $t_c = \xi/\gamma$ where $\xi = \gamma/I_s$. The use of $I_s$ in the calculation of $\xi$ is a necessary assumption, since otherwise we are not able to define the system by a unique characteristic length. In unit iteration interval ($\Delta i = 1$) the use of corresponding $N_s$ and $\Delta r$ values define $\Delta t = 1(T)$ for each $\Delta i = 1$. All computations are done well within the condition $\Delta t \ll t_c$, so that the finite time interval error is negligible [10]. Since we are using the finite difference numerical technique for these computations, therefore, there are no random errors associated with the data generated.

### *3.1. Homogeneous nucleation*

As noted earlier, time dependent nucleation rates violate the basic KJMA assumptions. As a first step, we therefore study the effect of increasing transient nucleation time $\tau$ on the Avrami exponent. Other conditions of the KJMA model are strictly followed. Thus, the transformation proceeds by homogeneous nucleation. The nuclei are of negligible size and grow in an isotropic manner at a constant rate.

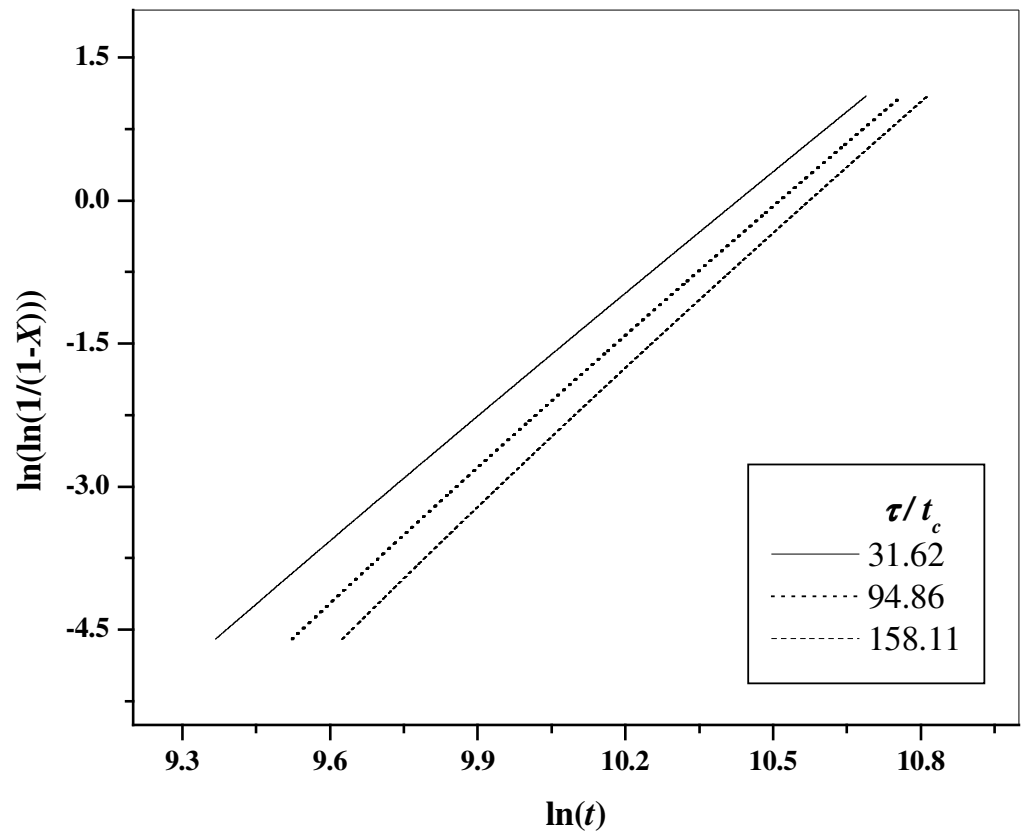

Fig. 1a

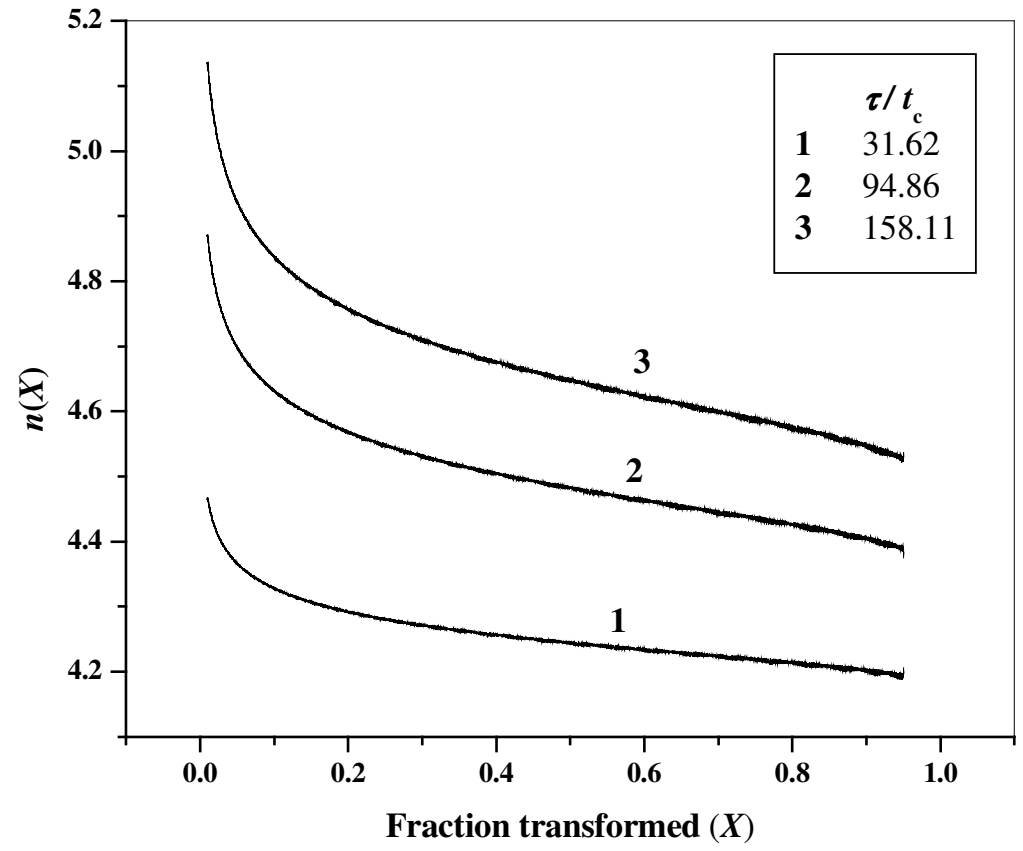

Fig. 1b

Fig. 1. a) $\ln(\ln(1/(1-X))$ versus $\ln(t)$ plots for systems at increasing dimensionless transient nucleation times ( $\tau/t_c$ ). b) Avrami exponents $(n)$ with increase in $X$ under homogeneous nucleation conditions. These are line plots (lines as guide to eye) prepared by smoothing the fluctuations in data points.

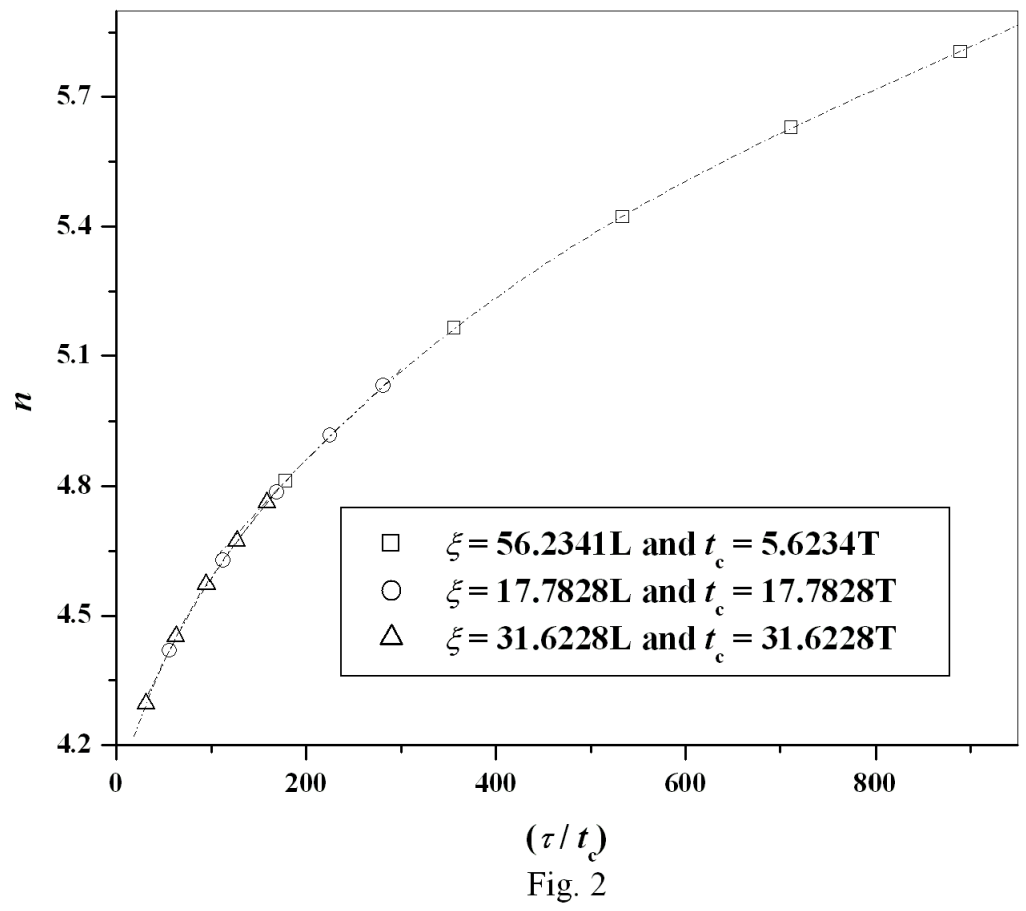

Fig. 2

Fig. 2. Average Avrami exponent ( $n$ ) versus dimensionless transient nucleation time $(=\tau/t_c)$ for systems defined by three different $\xi$ values (satisfying the condition $R_c \ll \xi$ ).

Fig. 1a shows the $\ln(\ln\left(1/(1-X)\right)$ versus $\ln(t)$ plots for systems defined by the same characteristic length but subjected to increasing transient nucleation times. The slopes of the best linear fits to $\ln(\ln\left(1/(1-X)\right)$ versus $\ln(t)$ plots give the Avrami exponent $(n)$ values. The correlation coefficient of the fits always satisfies the condition ($R$) > 0.99. For the sake of comparison across systems with different $\xi$, we represent the systems by their dimensionless transient nucleation times $(=\tau/t_c)$. As expected, we observe that increasing transient nucleation time's delays the time origin of transformation. In Fig. 1b we show the change in local Avrami exponents $\left[n(X)=\partial(\ln(\ln(1-X)))/\partial \ln t\right]$ with increase in the fraction of the product phase $(X)$ for the systems considered in Fig. 1a. Throughout the transformation $n(X)$ is always > 4. But, at the start of the transformation $n(X)$ is at a high value and then it decreases, initially at a rapid rate and later more gradually.

Fig. 2 gives the plots of $n$ values versus the dimensionless transient nucleation time $(=\tau/t_c)$ for systems defined by three different $\xi$ values (satisfying the condition $R_c \ll \xi$). We observe that the $n$ values for systems with same $\xi$ but different $\tau$ characteristics fall on a smooth non-linear curve, which fits a third-order polynomial function. However, there is no universality, since one single third order polynomial function does not fit the three different $\xi$ curves.

The aforementioned results correspond to the negligible nuclei size. It is already well known that including finite nuclei size can factor in bulk nanocrystallization kinetics for polymorphic transformations [10]. We now try to investigate the effect of finite nuclei size on the Avrami exponent scaling relations with respect to transient nucleation time. Fig. 3a shows Avrami exponent $n$ versus $\xi_c\left(=R_c/\xi\right)$ plots. Each plot is for system with different transient nucleation time. For effective comparison the transient nucleation times are divided by the corresponding characteristic time to yield dimensionless values $(\tau/t_c)$. We observe that for fixed value of the reduced transient nucleation time $(\tau/t_c)$ the Avrami exponent follows a linear scaling relation as $\xi_c \to 0$. In Fig. 3b we plot

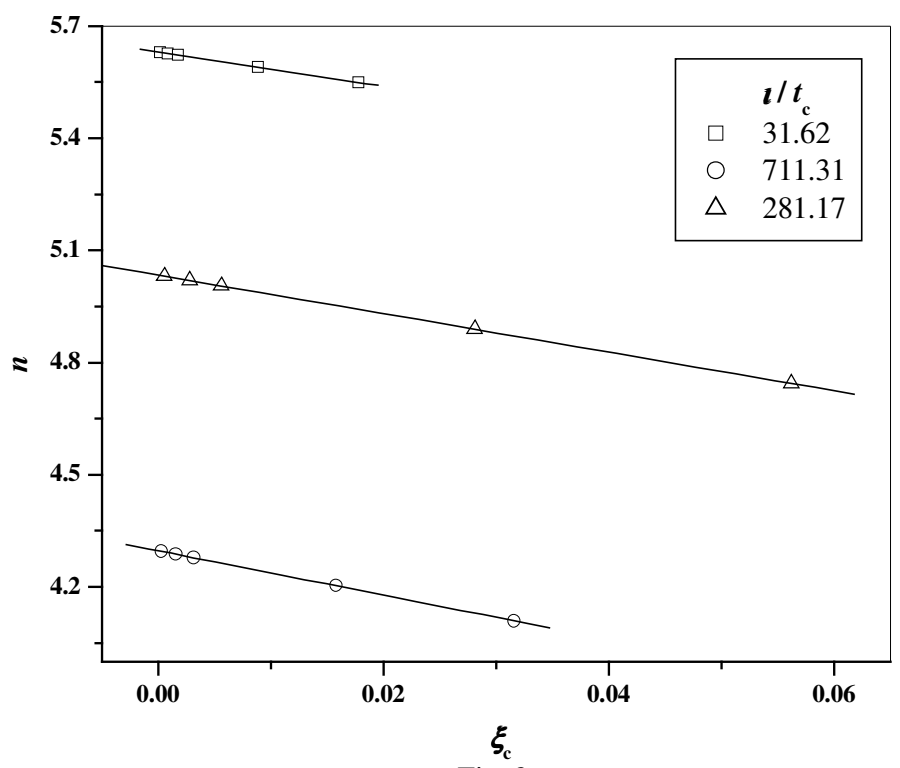

Fig. 3a

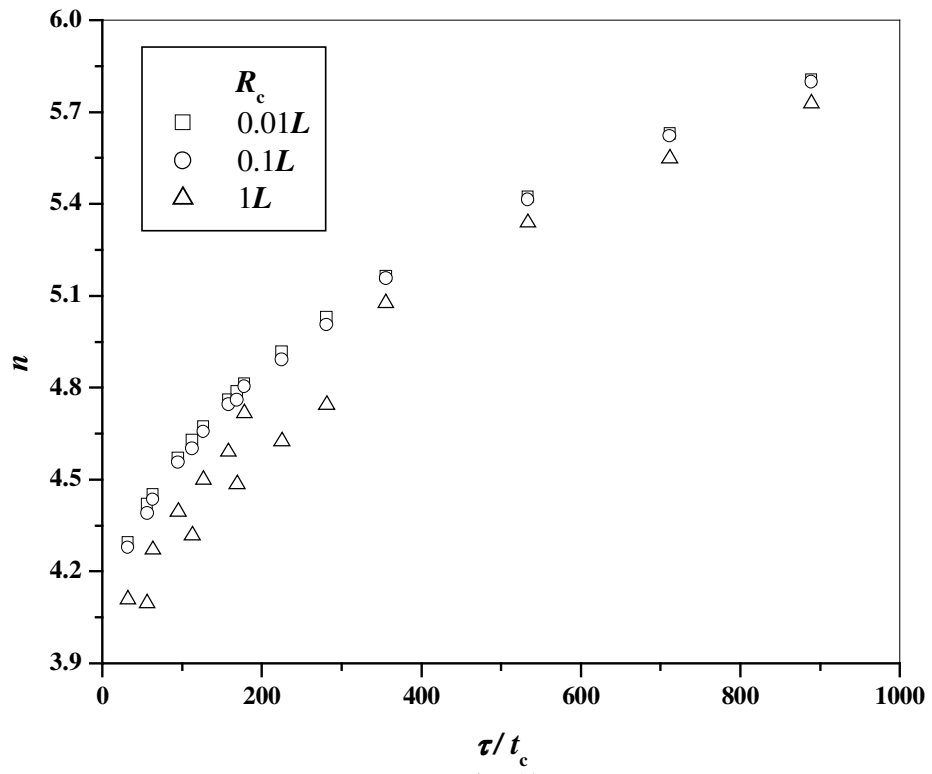

Fig. 3b

Fig. 3. a) Avrami exponent ($n$)versus $\xi_c\left(=R_c/\xi\right)$ plots at different dimensionless transient nucleation times $\left(=\tau/t_c\right)$. b) Avrami exponent $n$ versus dimensionless transient nucleation time$\left(\tau/t_c\right)$ at different $R_c$ values.

$n$ versus reduced transient nucleation time$\left(\tau/t_c\right)$. Plots are given for systems defined by different $R_c$ values. The plot for $R_c=0.1L$ is similar to that of $0.01L$. However, for $R_c=1L$ the plot becomes completely random in the earlier portion or shorter transient nucleation times relative to $t_c$.

*3.2 Heterogeneous nucleation*

In our previous work, we established that the Avrami exponents show a linear scaling relation with decrease in $R_c$ for a fixed $\theta$ value. Therefore, it is necessary that we first establish the relations for change in $n$ with $\theta$ keeping $R_c$ and $\xi$ fixed under constant nucleation rate conditions. We stress the importance of using $\xi$ instead of $\zeta=\left(V_0/N_t\right)^{1/D}$ where $N_t$ is the total number of potential heterogeneous nucleation sites in system. This is because in all of our computations the number $N_t$ is chosen such that it does not get exhausted and growth only stage is never encountered. To avoid any unaccounted effect, we fix the value of $\zeta$ for all the computations given here.

In Fig. 4a we plot $n$ versus $\theta$ for systems at different $\xi$ values. All plots are for systems with negligible critical nuclei ($R_c=0.01L$). For systems with different $\theta$ conditions having the same $\xi$ value, $n$ values follow a sigmoidal fit. Systems with the largest $\xi$ value and smallest $\xi_c$, show least variation. On comparing the three plots, we observe with increase in $\xi_c$, the variation in $n$ values increases. Fig. 4b shows $n$ versus $\theta$ plots for systems at different $\xi$ values and same $R_c = 1L$. The corresponding $\xi_c$ values are therefore larger by a factor of 100. The trend found in Fig. 4a is no more followed by systems with such large $\xi_c$ values. As $\xi_c$ increases from 0.0563 to 0.067 the fit to plots changes from sigmoid to Gaussian. Finally, for $\xi_c = 0.01$ we are only able to describe it by a fourth order polynomial fit. In Fig. 4c, keeping $\theta$ constant, we plot $n$ against $\xi_c$. The Avrami exponents scale linearly with $\xi_c$ when $\theta$ is held constant. For a given $\theta$, as we decrease $\xi_c$, the $n$ value tends to a fixed value for the negligible nuclei size condition.

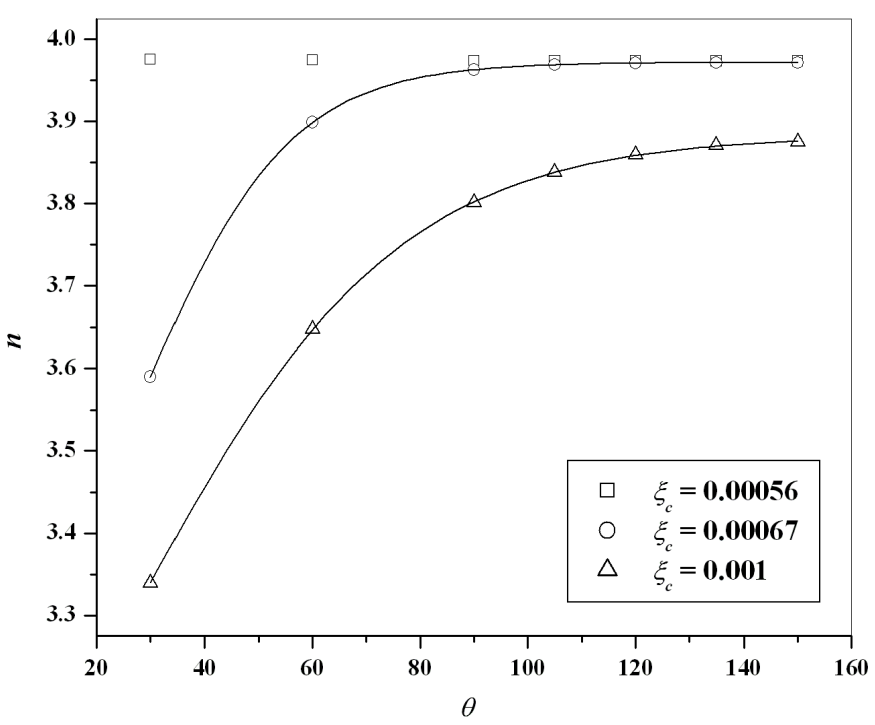

Fig. 4a

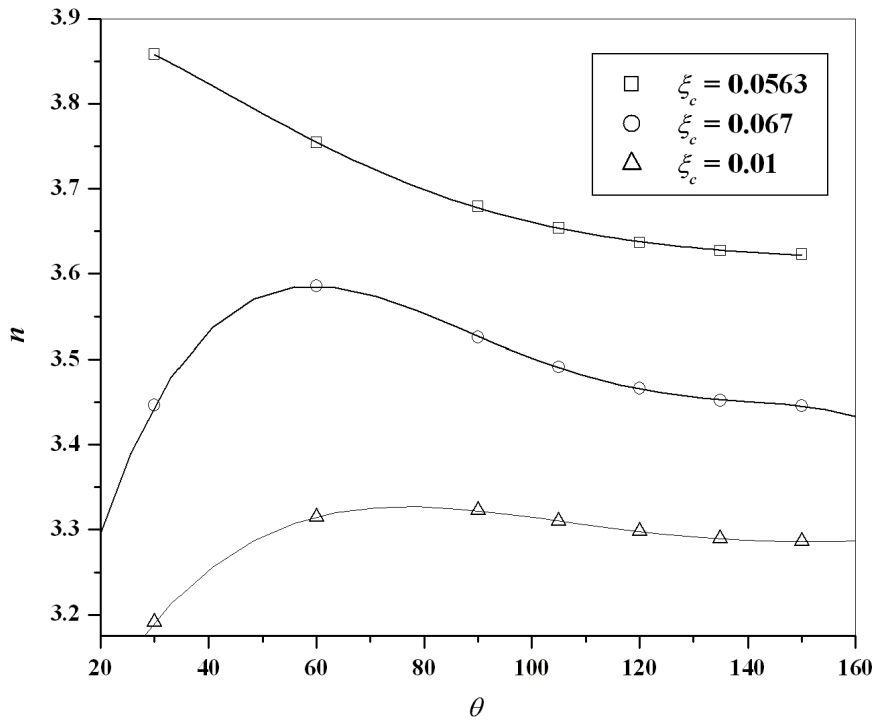

Fig. 4b

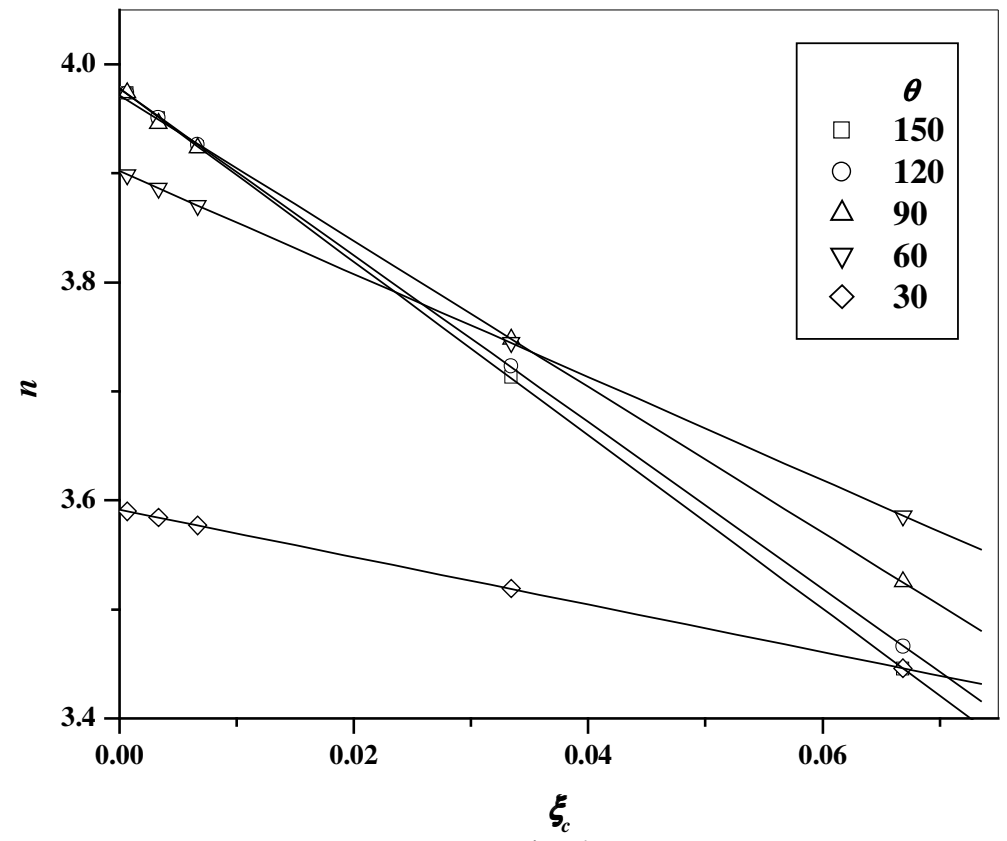

Fig. 4c

Fig. 4. Plots showing different aspects of transformation under heterogeneous but constant nucleation rate conditions. a) $n$ versus $\theta$ plots for systems at different $\xi$ values under $R_c \ll \xi$ conditions. Lines in different plots represent the best sigmoidal fits to the computed data. Correlation coefficients for all fits are >0.99. b) $n$ versus $\theta$ plots for systems at different $\xi$ values under $R_c < \xi$ conditions. Line in the plot $\xi_c = 0.0563$ represents the best sigmoidal fits to the computed data. Fourth order polynomial regression fits are obtained for other two $\xi_c$ values considered. Correlation coefficients for all fits are >0.99. c) Linear scaling relations shown by $n$ against $\xi_c$ plots at different $\theta$ values. Lines in different plots represent the best linear fits to the computed data. Correlation coefficients for all linear fits are equal to 1.

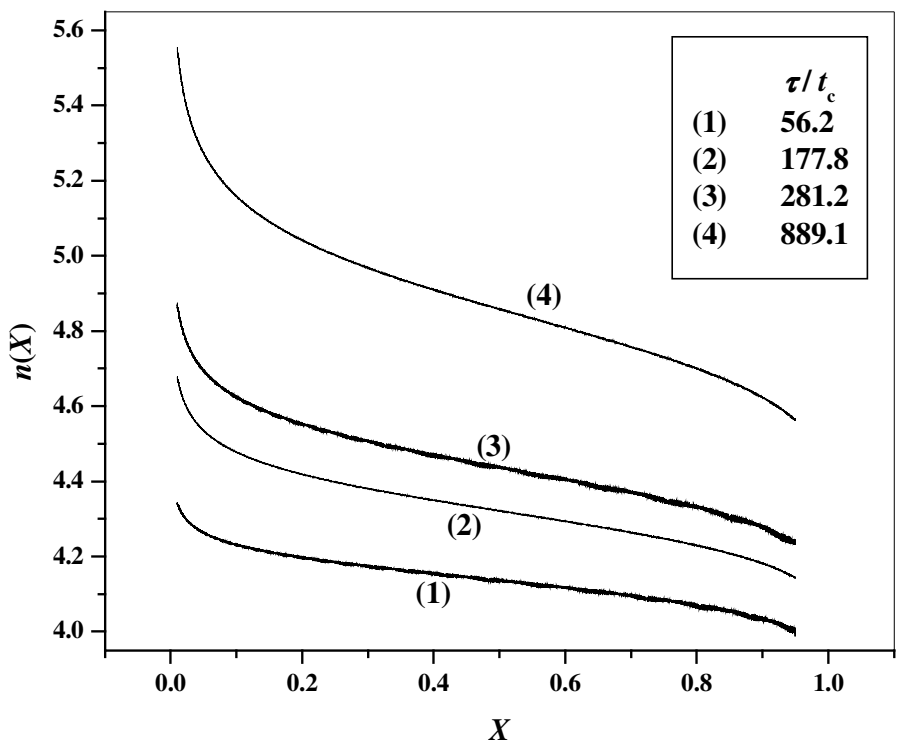

Fig. 5a

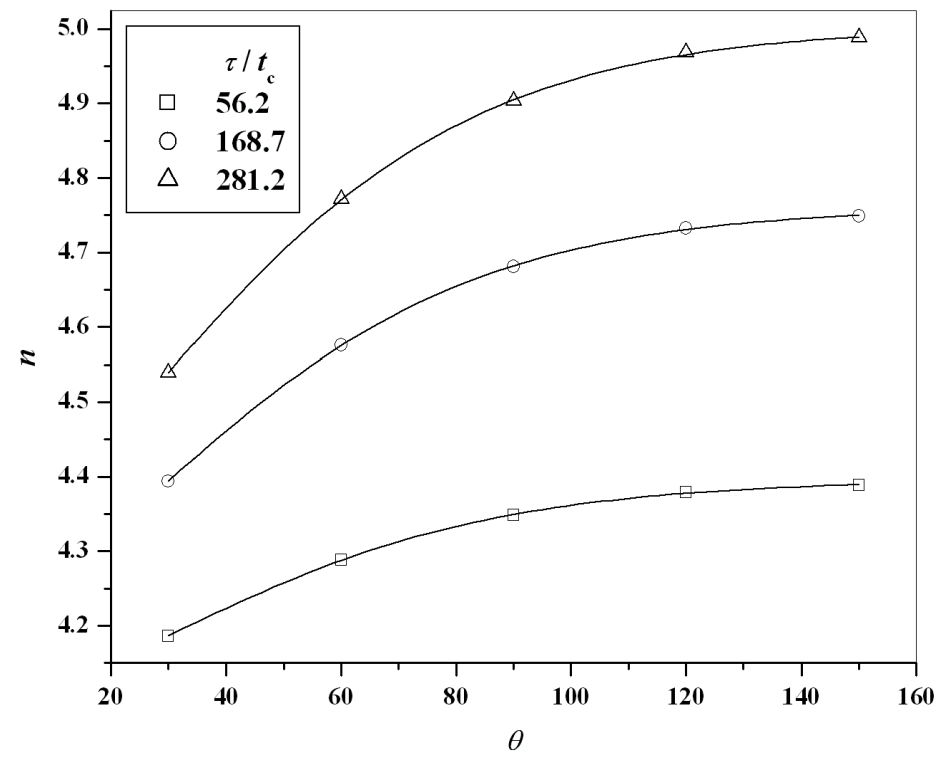

Fig. 5b

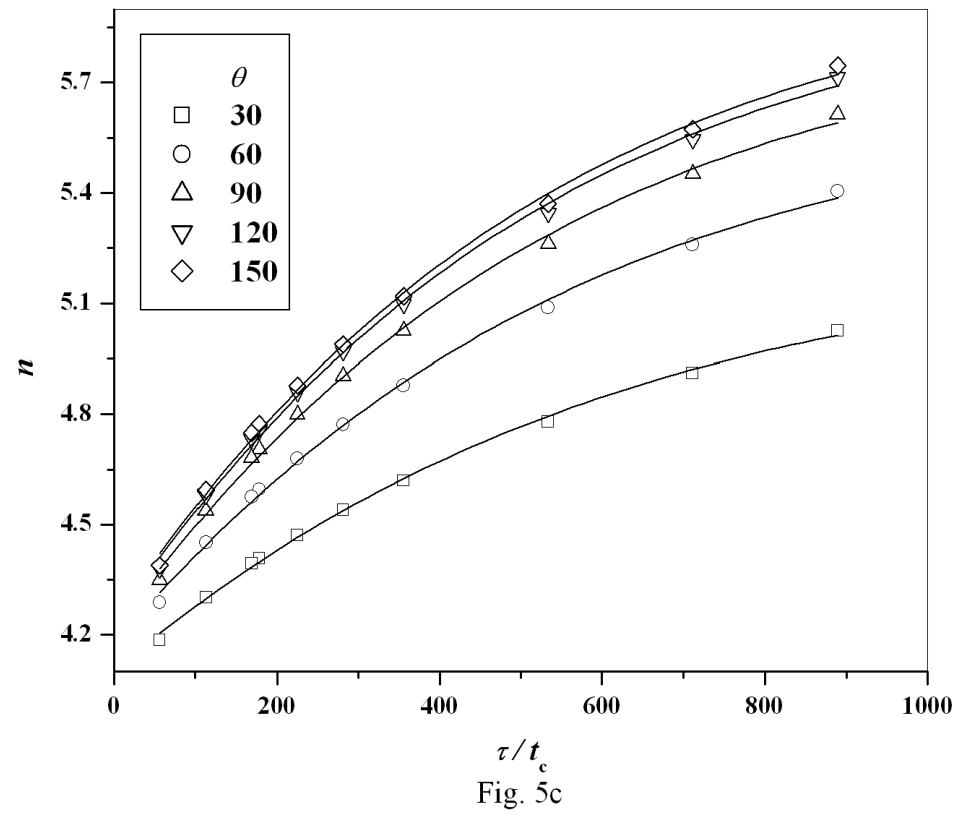

Fig. 5c

Fig. 5. Different facets of transformation kinetics under heterogeneous transient nucleation rate and $R_c \ll \xi$ conditions: a) Local Avrami exponent versus fraction transformed $(X)$ plots for systems (with contact angle $\theta = 30^\circ$) at different dimensionless $\tau/t_c$ values. These are line plots (lines as guide to eye) prepared by smoothing the fluctuations in data points. b) $n$ versus $\theta$ plots for systems at different dimensionless $\tau/t_c$ values. Lines in different plots represent the best sigmoidal fits to the computed data. Correlation coefficients for all fits are >0.99. c) $n$ versus $\tau/t_c$ plots for systems at different $\theta$ contact angle values. Lines in different plots represent the best sigmoidal fits to the computed data. Correlation coefficients for all fits are >0.99.

Now we present results for heterogeneous transient nucleation conditions. In Fig. 5, we consider systems with negligible nuclei radius conditions ($R_c \ll \xi$). Fig. 5a shows local Avrami exponent $n(X)$ versus fraction-transformed $(X)$ plots at different dimensionless $\tau/t_c$ values or transient nucleation times. As mentioned earlier, each plot is obtained by taking the derivative of the corresponding $\ln(\ln(1/(1-X))$ versus $\ln(t)$ plot against $X$. All plots in the figure are at $\theta = 30^\circ$, the minimum contact angle considered for heterogeneous nucleation cases in this study. We observe that the plots shift to higher range of $n$ values with increase in $\tau/t_c$ value. Otherwise, the nature of the curves does not change. As in systems transforming by homogeneous nucleation conditions, we take the slope of the best linear fit to $\ln(\ln(1/(1-X))$ versus $\ln(t)$ plot as the average Avrami exponent $(n)$. The correlation coefficients of such fit always satisfy $R > 0.99$. The average $n$ values obtained in this manner are plotted against $\theta$ in Fig. 5b. These plots constructed at increasing $\tau/t_c$ values are described by sigmoidal function fits. The other aspect of these plots (Fig. 5b) is given by $n$ against $\tau/t_c$ plots at different $\theta$ values in Fig. 5c. Perfect sigmoidal function fits describe the plots at all $\theta$ conditions.

Fig. 6 presents the results for the case of bulk nanocrystallization kinetics when the transformation is initiated by heterogeneous transient nucleation conditions. That is, now the nuclei size is finite in comparison to the characteristic length of the system $(R_c < \xi)$. Fig. 6a shows the local Avrami exponent $n(X)$ versus $X$ plots for this particular case. We observe two types of plots, depending on the $\tau/t_c$ values. While the latter portion of plots (1) and (2) are similar, the initial trends are different. Plots (3) and (4) are similar; however the average $n$ obtained is different. Fig. 6b depicts average Avrami exponent $(n)$ versus the contact angle $(\theta)$ plots at different $\tau/t_c$ values but the same $\xi_c\,(=0.0178)$. All plots show perfect sigmoidal fits. The next Fig. 6c shows the variation of average Avrami exponent $(n)$ with $\tau/t_c$. Each plot is at different $\xi_c$ and its value affects the range of $n$ value variation the most. Again all plots show perfect sigmoidal fits. Finally, Fig. 6d shows the variation of $n$ with $\xi_c$ at a fixed contact angle $(\theta = 30^\circ)$. Plots given are for different values of $\tau/t_c$. All plots show linear change with $\xi_c$.

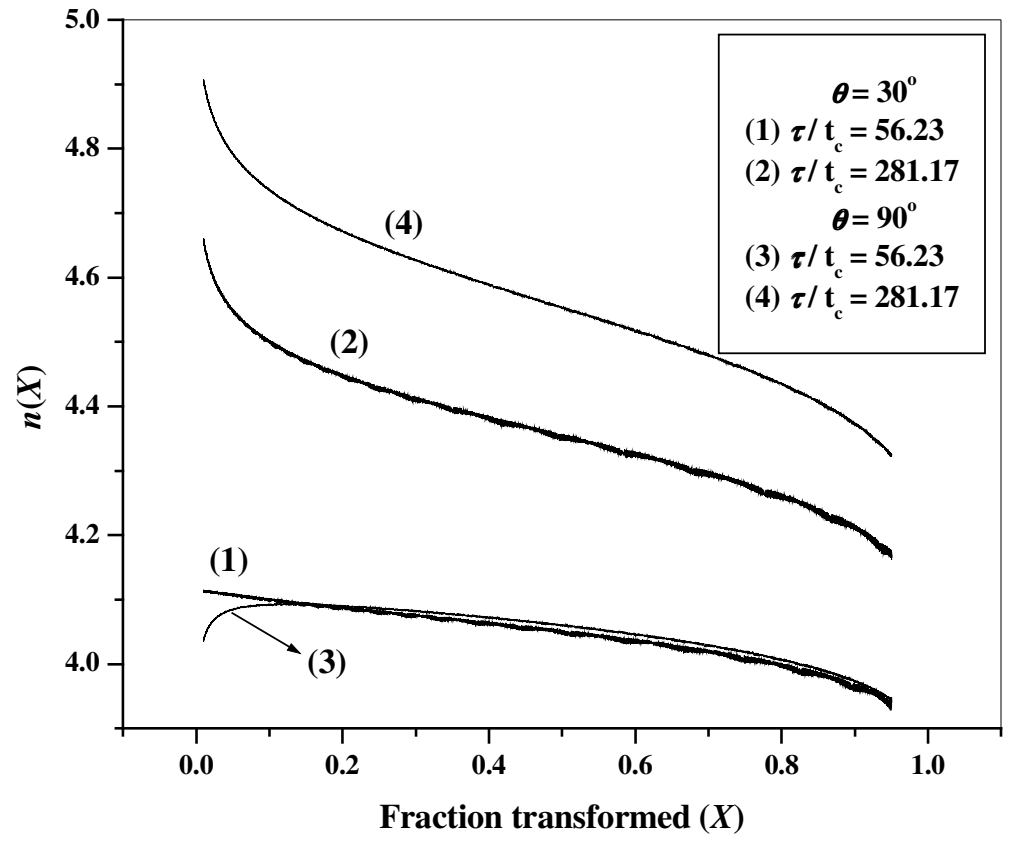

Fig. 6a

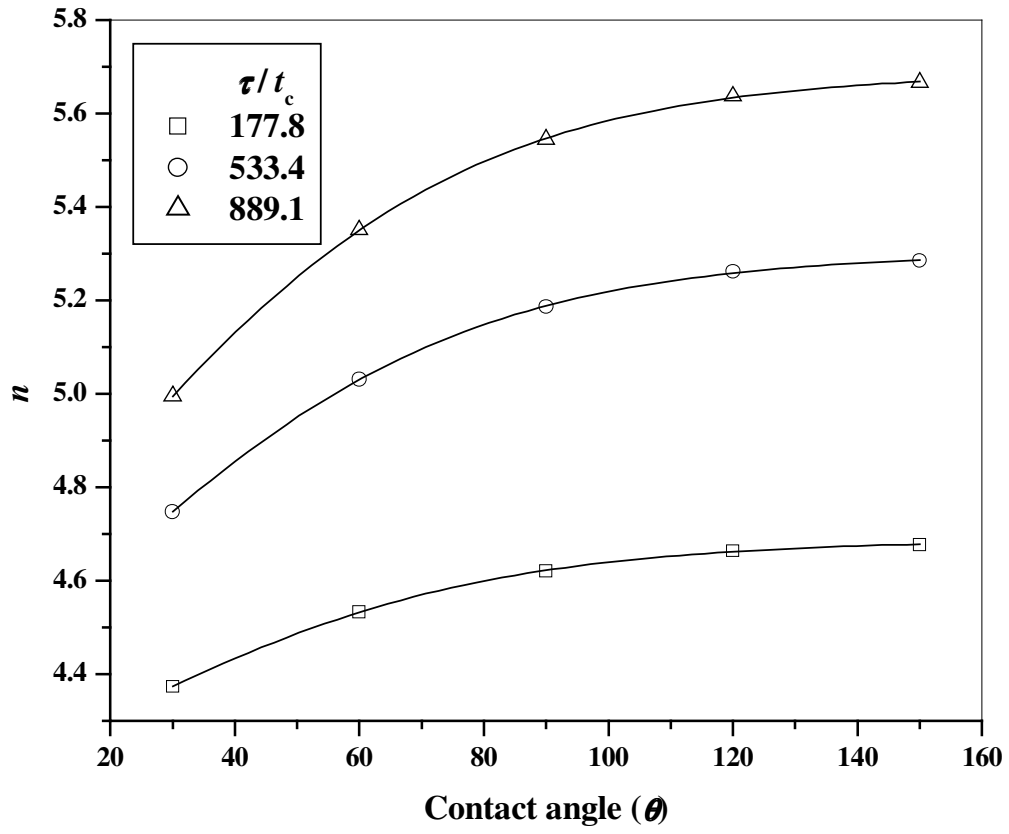

Fig. 6b

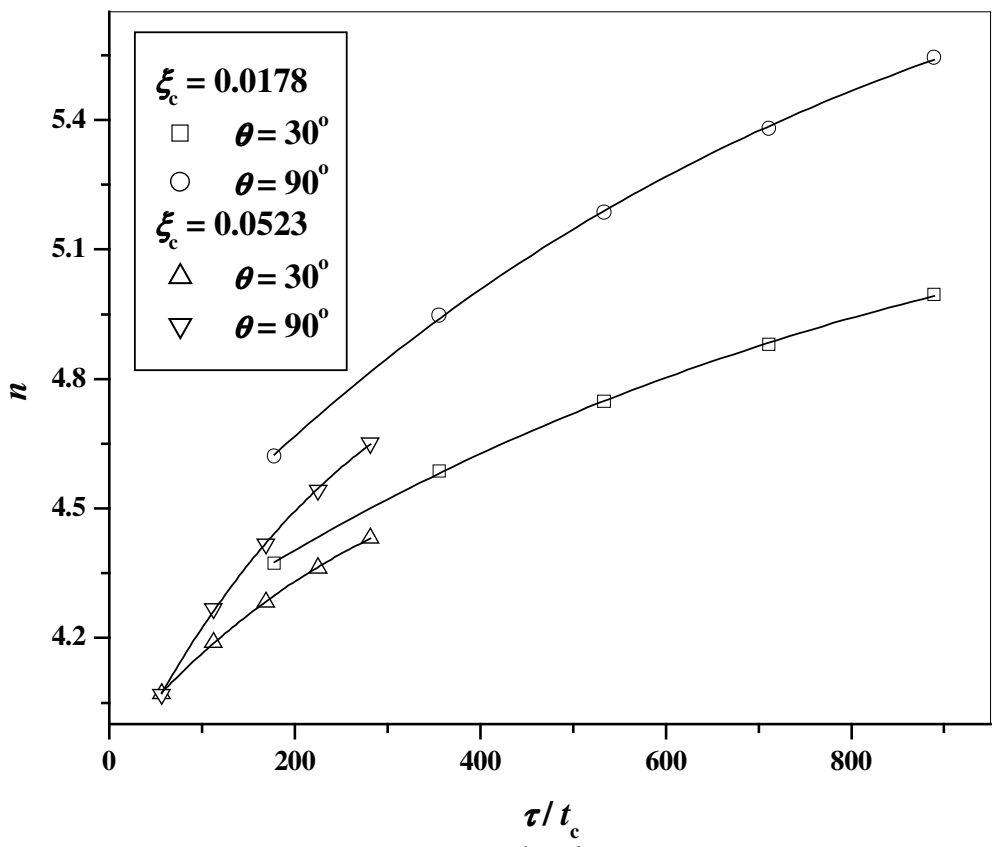

Fig. 6c

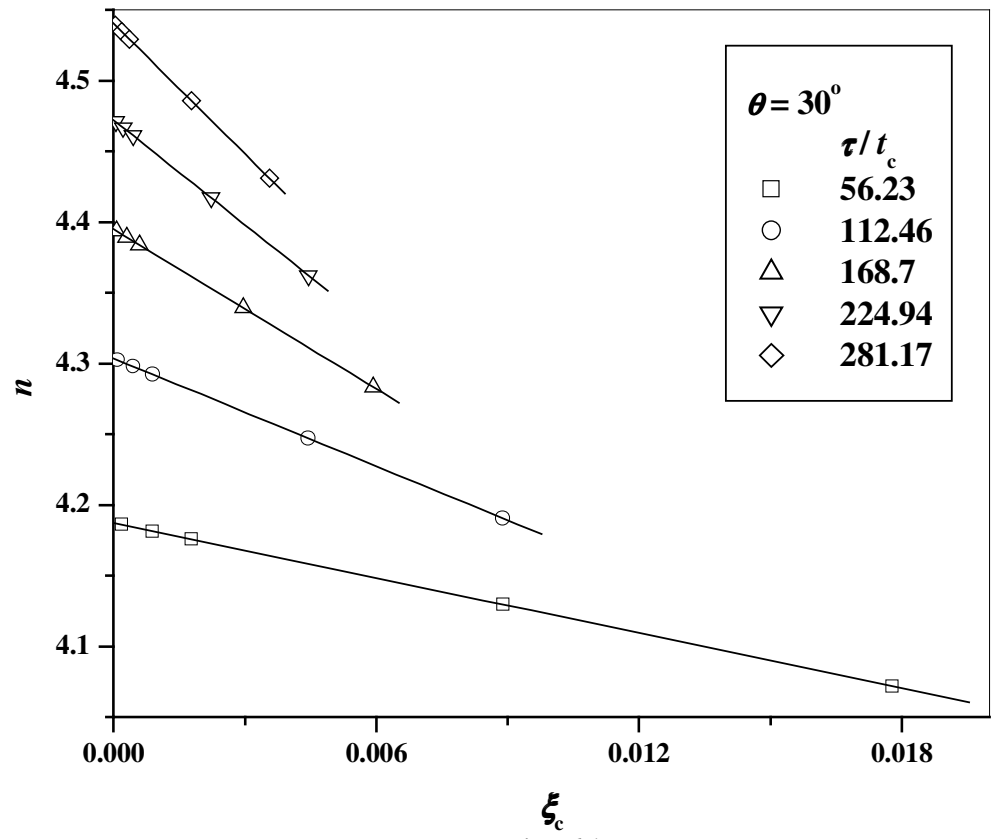

Fig. 6d

Fig. 6. Transformation kinetics under heterogeneous transient nucleation rate and $R_c < \xi$ conditions. a) Local Avrami exponent $(n)$ versus fraction transformed $(X)$ plots for systems (with contact angle $\theta = 30^\circ$ and $\theta = 90^\circ$) at different dimensionless $\tau/t_c$ values $(\xi_c = 0.0562)$. b) Plots of Avrami exponent $(n)$ against contact angle $\theta$ at different dimensionless $\tau/t_c$ values. All plots are at a fixed $R_c < \xi$ condition $(\xi_c = 0.0178)$. Lines in different plots represent the best sigmoidal fits to the computed data. Correlation coefficients for all fits are >0.99. c) Avrami exponent $(n)$ versus dimensionless $\tau/t_c$ plots for systems. Each plot is for a different contact angle $\theta$ and $\xi_c$ value combination. Lines in different plots represent the best sigmoidal fits to the computed data. Correlation coefficients for all fits are >0.99. d) Plots showing linear scaling of Avrami exponent $(n)$ with $\xi_c$. Correlation coefficients for all fits are >0.99.

## 4. Discussion

*4. 1 Variation of the local Avrami exponent*

Our calculations suggest for transformations initiated by homogeneous transient nucleation, the $n(X)$ value changes with time (Fig. 1b). The characteristic effect of transient nucleation is to increase the initial $n(X)$ value to > 4, depending on $\tau$. In the later stages it decreases and ultimately enters a steady state but always with a value > 4. This is similar to intermediate stages of experimental local Avrami exponent versus $X$ plots for crystallization kinetics of some metallic glass systems [31, 32]. Recently Zheng et al [33] conducted molecular dynamics (MD) simulations to study melt nucleation and growth process at atomistic scales in copper. After the incubation period, Avrami exponents gradually increased to values >4.

On the other hand heterogeneous nucleation leads to an exponent value < 4 [8, 15]. The variation of the local Avrami exponent $n(X)$ value versus $X$ for transformations initiated by heterogeneous nucleation at fixed contact angle values has been shown in reference [10]. On comparing Fig. 1b and Fig. 5a, we realize that the initially high value of $n(X)$, due to the magnitude of $\tau$ value, is diminished when the effect of heterogeneous nucleation is included. Also, in contrast to Fig. 1b, there is no steady $n$ value achieved in the later stages of the transformation. On further addition of the finite nuclei effect, the initial $n(X)$ value variation with $X$ is affected (Fig. 6a). The later stage transformation kinetics trend in fig. 6a remains similar to that seen in Fig. 5a. This has to be analyzed from the perspective of the negative deviation in the $n$ values when a system is subjected to a finite nuclei effect. As reported in reference [10], $n$ is at a low value at the start of the transformation and then increases to ultimately reach a steady value.

*4.2 Scaling relations for average Avrami exponent $(n)$*

In Fig. 2, we observe a near collapse of data points corresponding to systems with different $\xi$ values when the variation of $n$ with $\tau/t_c$ is considered. No such data collapse is observed when the effect of finite nuclei size is included (compare Fig. 2 and Fig. 3b). Finally, in Fig. 5c, we consider the effect of transient nucleation and heterogeneous nucleation together. As noted earlier, $n$ versus $\tau/t_c$ plots at different contact angles follow the similar sigmoidal functions. This suggests that for a given transient nucleation time we observe similar collapse of data points corresponding to systems defined by different $\xi$ values. Again, the data collapse observed in Fig. 5c is violated in Fig. 6c where, besides transient heterogeneous nucleation, systems also include the effect of nuclei of finite sizes.

We now compare the results of $n$ versus $\theta$ plots across systems with different features. The extent of deviation of the average value of $n$ from the universal value 4 is determined by the contact angle $\theta$, lesser the contact angle more the deviation (Fig. 4a). On evaluating Figs 4a and 4b, we find that for increasing $\xi_c$ value the plots deviate from the sigmoidal function. Finally, systems with large $\xi_c$ values may even show Gaussian or anomalous function dependence. Since, there is no transient nucleation, therefore, in all observations $3 < n < 4$. However, for systems with negligible $\xi_c$ values, transient nucleation does not affect the sigmoidal function dependence of $n$ versus $\theta$ plots. Thus, the plots in Fig. 5b are similar in nature to those in Fig. 4a. In contrast to Fig. 4b, in Fig. 6b we again observe sigmoidal dependence of $n$ versus $\theta$ plots for systems at different transient nucleation times initiated by nuclei of finite size $\left(R_c < \xi\right)$.

We now discuss the effect on the linear scaling relations found in reference [10] between the $n$ values of different systems, as their $\xi_c$ values tend to zero. When such plots are constructed at different transient nucleation times (Fig. 3a), they still give linear scaling relations. Further the slopes of these plots are also same, although with different intercepts. In contrast to this, in Fig. 4c and 6d, heterogeneous nucleation is also one of the factors. Although linear scaling relations are still followed, their slopes change for different systems.

**Conclusions**

We have investigated the effect of transient nucleation, transient heterogeneous nucleation as applied to bulk materials and materials with nano grain (or finite nuclei) sizes in the KJMA formalism. To delineate the effect of various factors we first consider separately transient and heterogeneous nucleation cases before taking them together. The local Avrami exponent change with fraction transformed has been described for each such factor. Non-linear (sigmoidal) scaling relations have been found for deviations from the universal Avrami exponent for transformations initiated by transient (homogeneous) nucleation. Limited universality is followed in terms of systems defined by different characteristic lengths. However, such universality is violated in different forms, as heterogeneous nucleation conditions are also included. Finally, even when transformation is by transient heterogeneous nucleation, we find that linear scaling relations are still followed between $n$ and $\xi_c$, although there is no universality.